\documentclass[twocolumn,floats,amsmath,amssymb,preprintnumbers,superscriptaddress,floatfix]{revtex4-1}

\usepackage{graphicx}
\usepackage{bm}
\usepackage{color}
\usepackage{hyperref}
\usepackage{amsmath}
\usepackage{amssymb}
\usepackage{mathtools}

 \usepackage{here}
 \usepackage[dvipsnames]{xcolor}

\newcommand{\figidx}[1]{(#1)}

\newcommand{\ie}{{\it{i.e.}}}
\newcommand{\eg}{{\it{e.g.}}}

\newcommand{\kT}{k_\text{B} T}
\newcommand{\enat}{e}

\usepackage{booktabs}
\AtBeginDocument{
	\heavyrulewidth=.08em
	\lightrulewidth=.05em
	\cmidrulewidth=.03em
	\belowrulesep=.65ex
	\belowbottomsep=0pt
	\aboverulesep=.4ex
	\abovetopsep=0pt
	\cmidrulesep=\doublerulesep
	\cmidrulekern=.5em
	\defaultaddspace=.5em
}

\usepackage{multirow}

\usepackage{siunitx}
\def\Eq{Eq.}
\def\Eqs{Eqs.}

\def\Fig{Fig.}
\def\Figs{Figs.}
\def\Tab{Table}

\extrafloats{100}

\begin{document}

\title{Tuning the Selective Permeability of Polydisperse Polymer Networks}

\author{Won Kyu Kim}
\email{wonkyukim@kias.re.kr}
\affiliation{Korea Institute for Advanced Study, Seoul 02455, Republic of Korea}

\author{Richard Chudoba}
\affiliation{Research Group for Simulations of Energy Materials, Helmholtz-Zentrum Berlin f\"ur Materialien und Energie, D-14109 Berlin, Germany}
\affiliation{Division of Theoretical Chemistry, Department of Chemistry, Lund University, P.O. Box 124, SE-22100 Lund, Sweden}

\author{Sebastian Milster}
\affiliation{Research Group for Simulations of Energy Materials, Helmholtz-Zentrum Berlin f\"ur Materialien und Energie, D-14109 Berlin, Germany}
\affiliation{Applied Theoretical Physics-Computational Physics, Physikalisches Institut, Albert-Ludwigs-Universit\"at Freiburg, D-79104 Freiburg, Germany}

\author{Rafael Roa}
\affiliation{Departamento de F\'{i}sica Aplicada I, Facultad de Ciencias, Universidad de M\'{a}laga, E-29071 M\'{a}laga, Spain}

\author{Matej Kandu\v{c}}
\affiliation{Jo\v{z}ef Stefan Institute, SI-1000 Ljubljana, Slovenia}

\author{Joachim Dzubiella}
\email{joachim.dzubiella@physik.uni-freiburg.de}
\affiliation{Research Group for Simulations of Energy Materials, Helmholtz-Zentrum Berlin f\"ur Materialien und Energie, D-14109 Berlin, Germany}
\affiliation{Applied Theoretical Physics-Computational Physics, Physikalisches Institut, Albert-Ludwigs-Universit\"at Freiburg, D-79104 Freiburg, Germany}
\affiliation{Cluster of Excellence livMatS @ FIT - Freiburg Center for Interactive Materials and Bioinspired Technologies, Albert-Ludwigs-Universit\"at Freiburg, D-79110 Freiburg, Germany}

\date{April 18, 2020}

\begin{abstract}
We study the permeability and selectivity (`permselectivity') of model membranes made of polydisperse polymer networks for molecular penetrant transport, using coarse-grained, implicit-solvent computer simulations. The permeability $\mathcal P$ is determined on the linear-response level using the solution--diffusion model, $\mathcal P = {\mathcal K}D_\text{in}$, \ie, by calculating the equilibrium penetrant partition ratio $\mathcal K$ and penetrant diffusivity $D_\text{in}$ inside the membrane. We vary two key parameters, namely the  monomer--monomer  interaction, which controls the degree of swelling and collapse of the network, and the monomer--penetrant  interaction, which tunes the penetrant uptake and microscopic energy landscape for diffusive transport. The results for the partition ratio $\mathcal K$ cover four orders of magnitude and are non-monotonic versus the parameters, which is well interpreted by a second-order virial expansion of the free energy of transferring one penetrant from bulk into the polymeric medium. We find that the penetrant diffusivity $D_\text{in}$ in the polydisperse networks, in contrast to highly ordered membrane structures, exhibits relatively simple exponential decays and obeys well-known free-volume and Kramers' escape scaling laws. The eventually resulting permeability $\mathcal P$ thus resembles the qualitative functional behavior (including maximization and minimization) of the partitioning. However, partitioning and diffusion are anti-correlated, yielding large quantitative cancellations, controlled and fine-tuned by the network density and interactions as rationalized by our scaling laws.  As a consequence, we finally demonstrate that even small changes of penetrant--network interactions, \eg, by half a $\kT$, modify the permselectivity of the membrane by almost one order of magnitude.
\end{abstract}


\maketitle

\section{Introduction}

Being a key transport property in materials science, the {\it permeability} of membranes has been excessively studied for more than a century~\cite{graham1866lv,finkelstein1987water,al1999one,permselectivity,venable2019molecular}.  The permeability determines the fundamental ability of functional solutes such as ions, ligands, proteins, and reactants to penetrate and be transported through dense but permeable membranes of various kinds. Membranes constitute typically quite crowded environments, are mostly polymer-based, and are ubiquitous in soft matter applications, materials science, and naturally in biological systems. In the latter, bio-hydrogels such as cytoskeletons, mucus gels, and the extracellular matrix (ECM) are complex molecular assemblies composed of hydrated polymer networks~\cite{shasby1982role,wingender1999bacterial,permselectivity,hay2013cell,witten2017particle,goodrich2018enhanced,fuhrmann2020diffusion} in cells. In general, they function as selectively permeable barriers for solutes to penetrate~\cite{witten2017particle}.
For instance, ECM constructs a selective barrier around the cells, thereby regulating transport of signaling molecules~\cite{taipale1997growth,dowd1999heparan,raines2000extracellular,garcia2003transport,thorne2008vivo,zhang2010role,theocharis2016extracellular,witten2017particle}.
Hence, the permeability of bio-hydrogels plays a decisive role in maintaining life.

Other important examples of polymer-network-based membranes can be found in functional soft matter composed of synthetic hydrogels, such as cross-linked poly(\textit{N}-isopropylacrylamide) (PNIPAM)~\cite{PNIPAM}.  Due to their thermoresponsiveness and relatively sharp volume transition they are widely used as representative and promising components in emerging material technologies for stimuli-responsive carrier particles, actuators, sensors, or responsive nanoreactors~\cite{Vriezema:2005fx,CarregalRomero:2010gp,Stuart:2010hu,Lu:2011bi,Renggli:2011if,Tanner:2011jf,Guan:2011eva,Herves:2012fp,Wu:2012bx,Gaitzsch:2015kr,Campisi:2016hl,Prieto:2016bj,Petrosko:2016he,Jia:2016cy}. In the latter, for instance, the hydrogel embeds nano-sized enzymes or metal nanoparticles catalyzing chemical reactions, which are ultimately controlled by the responsive membrane permeability~\cite{Stefano2015,Rafa2017,msde2020modeling}.  In general, responsive polymeric matrices can be expected to control permeation of (co)solute penetrants in a selective manner, modulated by external stimuli such as temperature, pH, and salinity. The tunable selectivity of the permeability (`permselectivity')~\cite{permselectivity} thus bears enormous potential for the development of `intelligent', programmable and adaptive membranes for diverse applications ranging from gas separation~\cite{robeson, chauhan, MOF, Lyd, Obliger, freeman1999basis,park2017}, water purification, and filtration~\cite{shannon:nature,geise2010water,geise2011,Menne2014,tansel,tan2018polyamide,hyk2018water} to dialysis and drug delivery~\cite{peppas1999, stamatialis}.

Typically, the permeability of dense membranes is quantified by the so-called solution--diffusion model on a linear-response level, via~\cite{yasuda1969permeability3,paul1976solution,robeson, Baker, gehrke, chauhan,al1999one,Thomas2001transport,Ulbricht2006,missner2009110,Baker2014,park2017,venable2019molecular} 
\begin{equation}\label{eq:P}
\mathcal{P} =  {\mathcal{K} D_\text{in},} 
\end{equation}
that is, it is the product of two key equilibrium quantities, namely the partitioning (partition ratio) $\mathcal{K}=c_\text{in}/c_{0}$,  simply given by the ratio between the penetrant concentrations inside the membrane $c_\text{in}$ and in the bulk $c_{0}$, and the diffusivity (diffusion coefficient) $D_\text{in}$  of the penetrants inside the membrane. The permeability of the bulk reference is thus equal to the free penetrant diffusivity in the bulk, \ie, $\mathcal{P}_0=D_0$. The elegance of \Eq~\ref{eq:P} is, that it is simply based on two intuitive and fundamental equilibrium properties of a medium, which should be presumably easy to access experimentally and theoretically tractable. 

However, there is a still growing number of theoretical studies pursuing a better understanding of partitioning~\cite{Ben, MOF, moncho2014ion, Lyd, Obliger,  adroher2015role,Erbas2016,Ben2, kim2017cosolute,monchoPCCP2018,kanduc2019aqueous} and diffusivity~\cite{Yasuda1968,yasuda1969permeability2,yasuda1969permeability3,robeson, Baker, gehrke, Torquato1991,masaro1999physical, Amsden1998,Chatterjee2011,Jiao2012,Spanner2013,Godec2014,Liasneuski2014,zhang2015particle,hansing2016nanoparticle,hansing2018particle,hansing2018hydrodynamic,hansing2018particleBJ,kim2019prl,Jha2011,Quesada-Perez2012,Kosovan2015,Kobayashi2016,Schmid2016,Srebnik2018,matej2018macro} in polymer-based membranes and hydrogels.  It is the complexity arising from diverse molecular interactions (\eg, excluded volume and attraction) and conformational structures (cross-linked, ordered, polydisperse) inside the membrane that renders the problem very challenging. In this context, for instance, we recently presented a simple coarse-grained (CG) simulation model of penetrant transport across a rigid immobile lattice-based membrane, pursuing a better comprehension of the permeability particularly in dense and attractive systems~\cite{kim2019prl}. Despite the simplicity of that model, we demonstrated a very intricate behavior of the permeability: the latter varied over many orders of magnitude, and could even be minimized or maximized by tailoring the potential energy landscape for the diffusing penetrants through small variations of membrane attraction, structure, and density. Supported by limiting scaling theories, we showed that the possible occurrence of extreme values is far from trivial, being evoked by a strong anti-correlation and substantial (orders of magnitude) cancellation between penetrant partitioning and diffusivity, especially in the case of dense and highly attractive membranes.

\begin{figure*}[ht!]
\centering
\includegraphics[width = 0.9\textwidth]{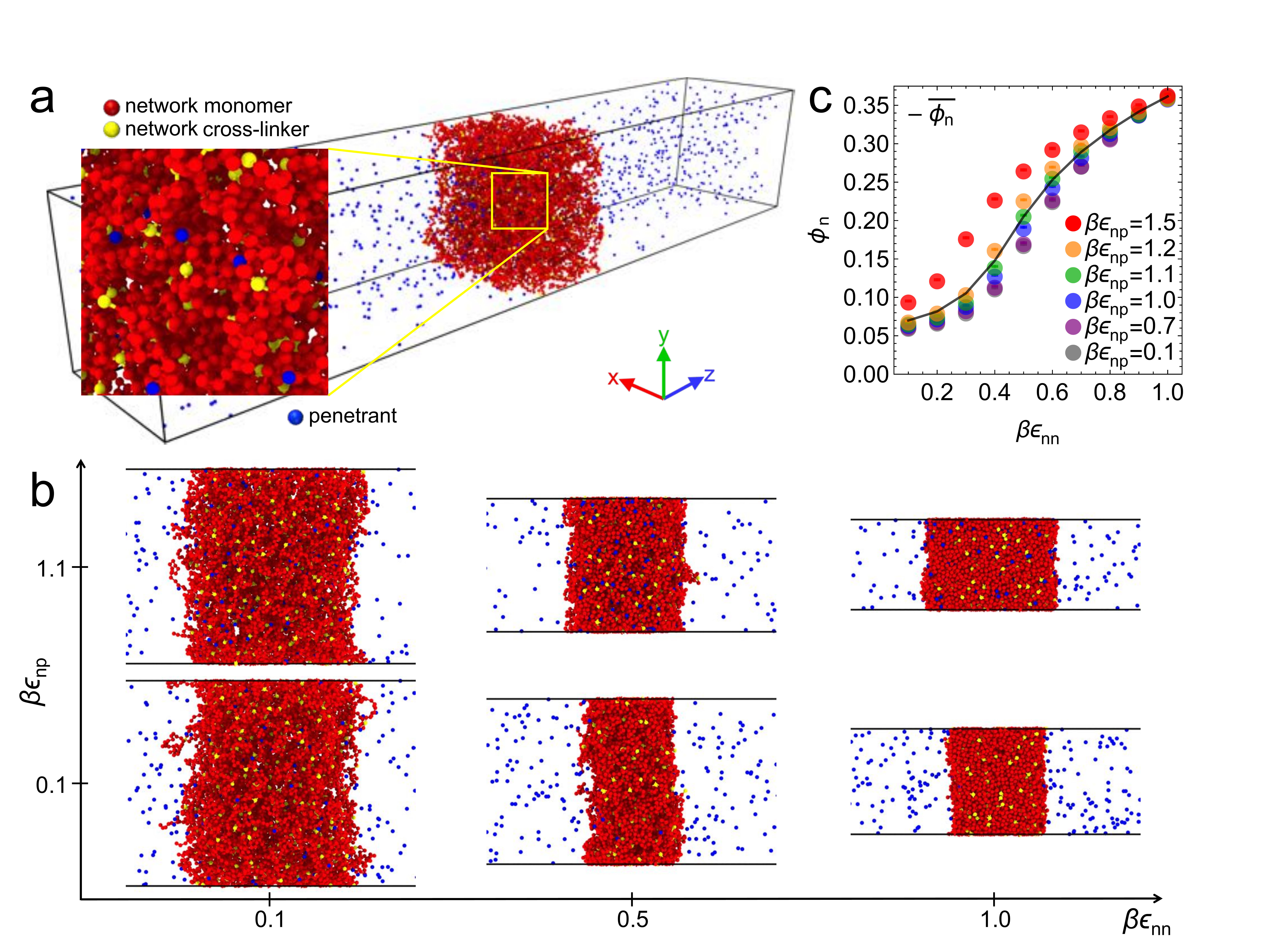}
\caption{
\figidx{a}~Simulation snapshot of the polydisperse tetra-functional polymer network in the swollen state with diffusive penetrants (blue).
 The polymer segments (red beads) are connected by tetra-functional cross-linkers (yellow beads) and have a random length distribution (see text in Methods). 
\figidx{b}~Various network conformations depending on $\epsilon_\text{nn}$ and $\epsilon_\text{np}$. The network collapses as the the network--network interaction parameter $\epsilon_\text{nn}$ increases (\ie, lowering the solvent quality to poor solvent conditions).
\figidx{c}~ The polymer network volume fractions $\phi_\text{n}$ vs. the solvent quality parameter $\epsilon_\text{nn}$ at different values of the network--penetrant parameters $\epsilon_\text{np}$ (see text for details). The solid line depicts the mean volume fraction $\overline{\phi_\text{n}}(\epsilon_\text{nn})$ interpolating between the averages over all simulated $\epsilon_\text{np}$ values.
}\label{fig:fig1}
\end{figure*}

In this work, we extend the previous study of a fixed, ordered membrane topology to a more complex and more realistic case of a membrane composed of fluctuating and cross-linked {\it polydisperse} polymers to study the transport of diffusive penetrants. For this, we consider a polydisperse tetra-functional network, i.e., each cross-linker connects four polymer strands, which have a polydisperse length distribution.  As similarly considered previously~\cite{kim2017cosolute,kim2019prl}, the system includes the network region and the bulk region, enabling a direct calculation of partitioning, diffusivity, and thus permeability.  We focus on two important control parameters: the polymer network density $\phi_\text{n}$ (volume fraction), tuned by internal interactions, and the interaction between the network monomers and the penetrants. We calculate the linear response permeability $\mathcal P$ according to eqn~(1) and systematically analyze and rationalize our findings by presenting semi-empirical scaling laws. Finally, we demonstrate how minute changes of the interactions can modify the permselectivity of the membrane substantially. 

\section{Methods}

\subsection{Simulation model}

\subsubsection{Network structure and setup}

We performed implicit-solvent Langevin dynamics computer simulations of the model membrane made of a polydisperse polymer network~\cite{higgs1988polydisperse,geissler1993scattering,glatting1995microscopic,soares2010mixture} including diffusive penetrants (see \Fig~\ref{fig:fig1}a), where each cross-linker connects four polymer chains but with different chain lengths. For the initial configuration of the network  we considered $4 \times 4 \times 4$ unit cells of a diamond cubic lattice, where $N_\text{x}=64\times8=512$ cross-linkers were located on the lattice points. The number of polymer monomers between the (closest neighboring) cross-linkers was randomly drawn from a uniform distribution between 2 and 18, thereby resulting in the polydisperse structure with an average chain length of 10 monomers, and a standard deviation of about 5. With the above construction we ended up with $N_\text{m} = 10364$ monomers in the network, yielding a cross-linker fraction of $f_\text{x} = 4.7~\%$.  This cross-linker fraction is in the range of typical experimental values for tetra-functional polymer networks, such as cross-linked PNIPAM hydrogels~\cite{Jha2011,Quesada-Perez2012,Kosovan2015,Kobayashi2016,Schmid2016}.

For initial equilibration, the membrane was placed in the middle of a simulation box of lateral lengths $L_x = L_y = 100 \sigma$ (with $\sigma$ defining the penetrant size and our length scale) and the longitudinal length $L_z = 300 \sigma$, with periodic boundary conditions in all three Cartesian directions.  The membrane was first equilibrated in the $NVT$ ensemble in the presence of the force-field described below. We then added $N_\text{p}=1000$ penetrant particles into the bulk region, and equilibrated the whole system. In the next step, the longitudinal box length $L_z$ was kept fixed, while $L_x $ and $L_y$ could adjust according to the $NpT$ ensemble with a given particle number $N = N_\text{m} + N_\text{x} + N_\text{p} = 11876$, pressure $p_x = p_y = p$, and temperature $T$.  The system was then finally allowed to equilibrate again before finally gathering statistics in the production runs.  

\begin{figure}[t]
\centering
\includegraphics[width = 0.47\textwidth]{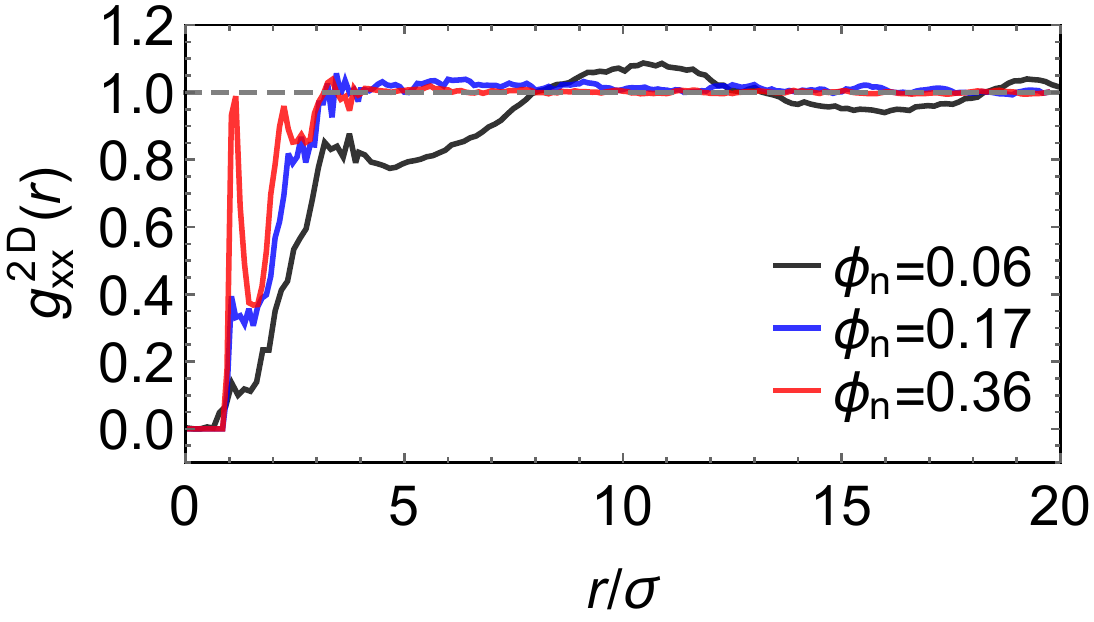}
\caption{The two-dimensional radial distribution function between the cross-linkers $g_\text{xx}^\text{2D}(r)$ for three different polymer volume fractions, from swollen (${\phi_\text{n}}=0.06$) to collapsed (${\phi_\text{n}}=0.36$) states. The network--penetrant interaction parameter is $\beta \epsilon_\text{np}=0.1$. The distribution function $g_\text{xx}^\text{2D}(r)$ is averaged over thin 2D membrane slabs in $xy$-directions, see ESI for details and more data.  }
\label{fig:fig_gx}
\end{figure}

Selected two-dimensional radial density distribution functions between the cross-linkers $g_\text{xx}^\text{2D}(r)$, shown in \Fig~\ref{fig:fig_gx}, demonstrate that the equilibration procedure leads to reasonable and homogeneous network structures (the $g_\text{xx}^\text{2D}(r)$ is averaged over thin two-dimensional membrane slabs in $xy$-directions, see the Electronic Supplementary Information (ESI) for details). Especially in the dense state, apart for some short-ranged packing effects, for $r \gtrsim 3 \sigma$ the system is very homogeneous.  For the swollen network, $g_\text{xx}^\text{2D}(r)$ reveals some more structure with a local peak in $3\sigma \lesssim r \lesssim 4\sigma$, reflecting short-range correlations between the crowded cross-linker regions, and a second peak close to the average chain length (\ie, average mesh size in the swollen case) of $r\simeq 10\sigma$. Changing the network--penetrant interaction affects these distributions only slightly in the dense systems, while in the swollen case some homogenization is observed for large attractions between the network and the penetrants (see \Fig~S1 in ESI). 
  
As described in our previous studies~\cite{kim2017cosolute,kim2019prl}, we employed the LAMMPS software~\cite{Plimpton1995} with the stochastic Langevin integrator in the $NpT$ ensemble. To maintain fixed pressure, we used the Berendsen anisotropic barostat~\cite{Berendsen1984}. The iteration time step $\delta \tau~=~5 \times 10^{-3} \tau$ was used with the time units $\tau =~\sqrt{ {m \sigma^2} / {\kT} }$, where $m$ is the unit mass, and $\kT=1/\beta$ is the thermal energy. The friction coefficient $\gamma$ was chosen to have the momentum relaxation time $\tau_{\gamma}=~m/\gamma = \tau$, so that the free penetrants' motion becomes diffusive after $200$ time steps.  The value of the lateral pressure was chosen to be $p~=~6.5853 \times 10^{-4}~\kT/\sigma^3 \approx 1$~bar.  The pressure relaxation time $\tau_{p}$ and bulk modulus $K_\text{b}$ for the barostat were carefully chosen in the range of $1\leq \tau_{p}/\tau \leq2$ and $1 \leq K_\text{b}/p \leq10$, respectively, depending on the interaction parameters. After an equilibration time of $1.5\times10^5~\tau$, we performed the production simulations typically up to $10^7 \delta \tau = 5 \times 10^4~\tau$.  As the finite network membrane is connected to the large bulk region of solute penetrants, the simulations are effectively isobaric/semi-grand canonical in the sense that the penetrants can always equilibrate their partitioning between the large bulk and the (responsive) membrane, subjected to the constant lateral pressure.

\subsubsection{Force field}

For the non-bonded interactions, all particles (i.e., monomer, crosslinker, penetrant beads) interact via the generic Lennard-Jones (LJ) potential $U_\text{LJ}^{ij}$ for $i, j =$ n or p, where n denotes the network particles (polymer monomers and cross-linkers), and p denotes the penetrant. The strength $\epsilon_\text{pp} = 0.1~\kT$ of the LJ potential $U_\text{LJ}^\text{pp}$ is fixed such that the penetrants are overall repulsive~\cite{kim2017cosolute,kim2019prl} (see also the positive second virial coefficient of the LJ interaction plotted in \Fig~S2a in ESI, and the following sections for details of the virial coefficients). In this work we vary two interaction parameters, the network--network interaction $\epsilon_\text{nn}$, and the network--penetrant interaction $\epsilon_\text{np}$, between $0.1$ and $1.5$~$\kT$.  The intra-network interaction $\epsilon_\text{nn}$ is interpreted as a measure of solvent quality~\cite{Heyda2013,kim2017cosolute,kim2019prl}, thereby controlling the network volume fraction $\phi_\text{n}$. As discussed in previous works~\cite{Heyda2013,kim2017cosolute,kim2019prl}, small/high $\epsilon_\text{nn}$ corresponds to good/poor solvent leading to a small/high volume fraction, respectively. The network--penetrant interaction $\epsilon_\text{np}$ governs the strength of the attraction between the polymers and the penetrants.  

For the bonded interactions of the (semi-flexible) polymers we employed harmonic stretching (bonds) and bending (angles) potentials~\cite{kim2017cosolute}.  The bonded polymer parameters were determined via coarse-graining from explicit-water, all-atom simulation results of cross-linked PNIPAM chains, utilizing a force-field from our group's work~\cite{milster2019cross}.  Since the cross-linker  (x) connects monomers (m) of four polymer chains, the network is tetra-functional, and in addition to the m-m-m bending, there are six bending potentials for the m-x-m arrangement.  Therefore, we have nine different bonding (7 bending (angles), 2 stretching (bonds)) potentials in total and we determined eighteen bond parameters $K_r^{ij}$, $r_0^{ij}$, $K_\theta^{ijk}$, and $\theta_0^{ijk}$ by fitting harmonic potentials to the free energies obtained from the all-atom simulations.  The details of all the bonded interactions, that is, their calculation from the all-atom (explicit-water) simulations of PNIPAM and their final definition, can be found in ESI.

\subsection{Analysis}

The partition ratio, $\mathcal{K} = c_\text{in} / c_{0}$, was computed by counting and averaging the equilibrium number density of penetrants inside the network and bulk, as similarly done in our previous works~\cite{kim2017cosolute,kim2019prl,msde2020modeling}: we carefully divided the simulation box into three regions  (inner membrane, membrane surface, and bulk) to sample the concentrations without any surface effects (due to the finite membrane width). See \Fig~S3 ESI for details. 

To calculate the penetrant diffusivity in the network, $D_\text{in}$, we generated 20 simulation boxes of diamond unit cells of the polydisperse tetra-functional networks including the penetrants for each parameter set of $\epsilon_\text{nn}$ and $\epsilon_\text{np}$, and we performed additional simulations of these periodic cells (see \Fig~S4 in ESI).
To determine the cell size and the number of the penetrants in the cell, we used the equilibrium values of the penetrant density and the polymer density obtained from the main simulation data.
We computed the mean-squared-displacement (MSD) of the penetrants in the networks, averaged over time and particles~\cite{shin2017elasticity}, as shown in supplementary \Fig~S5 (upper panels), within the dimensionless simulation time range from $t = 100$ to $t = 1000$ to obtain diffusivity via $\text{MSD} = 6 D_\text{in} t$, ensuring the normal diffusion~\cite{shin2017elasticity}, which fulfills $\alpha = {{\rm{d}} \ln \text{MSD} \over {\rm{d}} \ln {t} }= 1$ in \Fig~S5 (lower panels).

\section{Results and discussion}

\subsection{Network density response to solvent quality and penetrants}

\begin{figure*}[t]
\centering
\includegraphics[width = 0.95\textwidth]{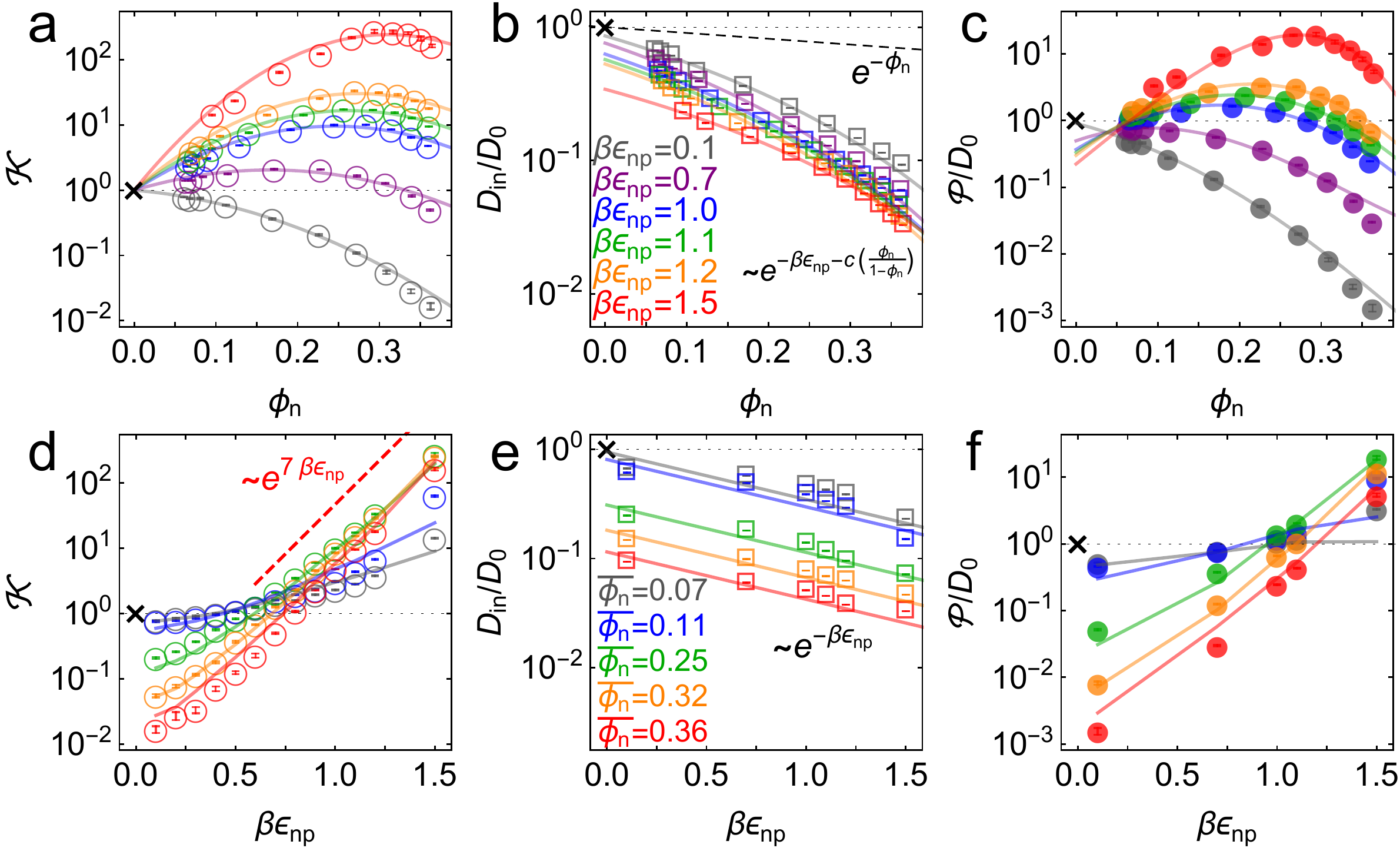}
\caption{Simulation results (symbols) and theoretical fits (solid lines) are shown. 
\figidx{a}~Partitioning $\mathcal{K}$ as a function of the network volume fraction for different values of the network--penetrant interaction $\epsilon_\text{np}$.
The solid lines are the fits from virial expansion \Eq~\eqref{eq:b2b3} (see text and \Tab~S1 in ESI for details).
\figidx{b}~Penetrant diffusion in the polymer network $D_\text{in}$ as a function of $\phi_\text{n}$.
The solid lines are the fits (also depicted in \figidx{e}) from the free-volume approach \Eq~\eqref{eq:D_scale} (see \Tab~S2 in ESI for the details).
The dashed line is $\enat^{-\phi_\text{n}}$ as a reference.
\figidx{c}~Permeability $\mathcal{P}$ as a function of $\phi_\text{n}$.
The solid lines are the predictions from \Eq~\eqref{eq:perm} using the fitting values from (a) and (b). 
\figidx{d}~Partitioning $\mathcal{K}$ as a function of $\epsilon_\text{np}$.
The solid lines are the prediction according to the virial expression \Eq~\eqref{eq:b2b3} using the $B_2^\text{np}$ and $B_3^\text{nnp}$ values from the fit in (a) and the mean density ${\phi_\text{n}}=\overline{\phi_\text{n}}$.   The data scales roughly as $\mathcal{K} \sim \enat^{7 \beta\epsilon_\text{np}}$ for dense and attractive polymers (red dashed line). 
\figidx{e}~Penetrant diffusion $D_\text{in}$ as a function of $\epsilon_\text{np}$.
The solid lines are the fits according to the exponential (Kramers') scaling $D_\text{in}/D_\text{0} \sim \enat^{-\beta\epsilon_\text{np}}$ from \Eq~\eqref{eq:D_scale}.
\figidx{f}~Permeability $\mathcal{P}$ as a function of $\epsilon_\text{np}$.
The solid lines are the predictions from \Eq~\eqref{eq:perm} using the fitting values from (a) (see also panels (d) and (e)).
The cross symbols in each panel are corresponding reference values in the bulk.}
\label{fig:fig2}
\end{figure*}

Six representative simulation snapshots of the total system are shown in \Fig~\ref{fig:fig1}b for different values of the solvent quality parameter $\epsilon_\text{nn}$ and the network--penetrant interaction parameter $\epsilon_\text{np}$.  The most swollen state is shown by the lower left snapshot, whereas the most compact state is depicted by the upper right snapshot. The polymer network collapses due to strong network--network attractions $\epsilon_\text{nn}$ (poor solvent), otherwise it swells (good solvent).
In addition, upon changing the network--penetrant interaction, particularly at the intermediate solvent quality ($\epsilon_\text{nn}=0.5~\kT$), we note that the larger the attraction $\epsilon_\text{np}$, the more packed is the network (lower volume). This is due to bridging effects of highly attractive penetrant, contracting the network to maximize favorable interaction contacts~\cite{kim2017cosolute}. See \Fig~S6 in ESI for details of the network volume change depending on the interactions.

The global effects of the two interaction parameters are summarized in \Fig~\ref{fig:fig1}c, which depicts the network volume fraction $\phi_\text{n}$, \ie, the ratio of the volume occupied by the polymers to the entire network volume, as a function of the solvent quality parameter $\epsilon_\text{nn}$ for different values of the network--penetrant interaction $\epsilon_\text{np}$. The network undergoes a typical collapse transition as $\epsilon_\text{nn}$ increases at small and intermediate values of $\epsilon_\text{np}$, while the transition becomes more gradual when the attraction is very high ($\epsilon_\text{np} = 1.5~\kT$). This is probably due to local monomer clustering and network homogeneity, cf. \Fig~S1,  smoothening the transition.  Note that in \Fig~\ref{fig:fig1}c we also depict the mean volume fraction $\overline{\phi_\text{n}}(\epsilon_\text{nn})$, which is the average over all simulated $\epsilon_\text{np}$.

\begin{table*}[t]
\centering
\caption{\label{Tab:B2}Virial coefficients $B_2^\text{np}$ and $B_3^\text{nnp}$ in \Eq~\eqref{eq:b2b3} obtained as fitting parameters in \Fig~2a are shown for different values of $\epsilon_\text{np}$. The exact values of $B_2^\text{LJ}$ and $B_3^\text{LJ}$ for LJ potential are shown for comparison. The fitting parameter values $b$ and $c$ in the free-volume scaling theory \Eq~\eqref{eq:gfree} (see also \Fig~\ref{fig:fig2}b) are shown. See ESI for details.}
\sisetup{table-format = -2.2(2)}
\begin{tabular}{lSSSSSS}
\toprule
$\beta \epsilon_\text{np}$ & 0.1 & 0.7 & 1.0 & 1.1 & 1.2 & 1.5 \\
\midrule
$B_2^\text{np}/\sigma^3$ & 0.63(8) & -2.25(17) & -4.82(19) & -5.61(16) & -6.52(21) & -9.26(34)\\
$B_2^\text{LJ}/\sigma^3$ & 0.97 & -2.77 & -5.32 & -6.27 & -7.29 & -10.75\\
$B_3^\text{nnp}/\sigma^6$ & 3.85(35) & 4.74(44) & 6.85(45) & 7.46(38) & 8.31(48) & 10.41(74)\\
$B_3^\text{LJ}/\sigma^6$ & 1.25 & 2.46 & 1.89 & 0.55 & -1.92 & -22.07\\
\midrule
$b$ & 0.86(2) & 0.76(2) & 0.63(2) & 0.57(1) & 0.53(2) & 0.34(1)\\
$c$ & 4.17(19) & 4.81(16) & 4.77(19) & 4.69(18) & 4.79(20) & 3.87(9)\\
\bottomrule
\end{tabular}
\end{table*}

\subsection{Penetrant partitioning, diffusivity, and permeability}

\subsubsection{Partitioning}

In \Fig~\ref{fig:fig2} we show the partitioning $\mathcal{K}$, the penetrant diffusion inside the network $D_\text{in}$, and the permeability $\mathcal P$, as a function of the network volume fraction (a--c) and the network--penetrant interaction parameter (d--e).

The partitioning as a function of the network volume fraction, $\mathcal{K} ( \phi_\text{n} )$, exhibits diverse behavior, ranging over four orders of magnitude depending on the  interactions, as shown in \Fig~\ref{fig:fig2}a.  For low network--penetrant interaction parameters $\epsilon_\text{np}$, $\mathcal{K}$ is monotonically decreasing with increasing network density, since the essentially repelled penetrants are excluded by highly packed polymers (see the second virial coefficient of the LJ system shown in \Fig~S2a in ESI). For higher values of the LJ potential depth $\epsilon_\text{np}$, the penetrants are increasingly more attracted to the network. The partitioning $\mathcal{K}$, however, becomes non-monotonic and reaches a maximum around $\phi_\text{n} \simeq 0.3$. This partitioning maximization is due to the volume exclusion of the penetrants, which wins over the attraction at high densities~\cite{monchoPCCP2018, kim2019prl}.


The cross-over from penetrant exclusion to enrichment for increasing $\epsilon_\text{np}$ at fixed polymer density $\phi_\text{n}$ becomes obvious in \Fig~\ref{fig:fig2}d, where we plot $\mathcal{K}(\epsilon_\text{np})$.  At around $\beta \epsilon_\text{np} \simeq 0.5 - 0.7$ (depending in detail on polymer density) the attraction outvalues the steric obstruction and penetrants are on average preferentially adsorbed than being in bulk, \ie, $\mathcal K>1$.  We also observe that the partitioning $\mathcal{K}(\epsilon_\text{np})$ exhibits roughly an exponential increase with larger slope as $\phi_\text{n}$ increases. The exponential increase of the partitioning is also found in ordered membranes~\cite{kim2019prl}, reflecting that the overall scaling behavior of partitioning (upon changing the interactions) is rather insensitive to the regularity of the network.
For dense and attractive polymer networks, we empirically find that $\mathcal{K} \sim \enat^{7 \beta\epsilon_\text{np}}$, as depicted in \Fig~\ref{fig:fig2}d. The prefactor 7 reflects the total mean attraction in the dense systems, where the potential wells of many attractive monomers densely overlap. 

In order to gain more theoretical insight and develop an analytical framework for describing the data, we perform a virial expansion of the transfer free energy $\beta\Delta G \simeq~2 B_2^\text{np} {\phi_\text{n}}/{v_0} + \frac{3}{2} B_3^\text{nnp} (\phi_\text{n}/{v_0})^2$, and apply it to the partition coefficient $\mathcal K~=~\exp(-\beta \Delta G)$~\cite{kim2019prl}, as 
\begin{eqnarray}\label{eq:b2b3}
\mathcal{K} = \exp \left[ -2 B_2^\text{np} \frac{\phi_\text{n}}{v_0} - \frac{3}{2} B_3^\text{nnp} \left( \frac{\phi_\text{n}}{v_0} \right)^2 \right],
\end{eqnarray}
where $B_2^\text{np}$ is the second virial coefficient, $B_3^\text{nnp}$ the third virial coefficient, and $v_0 = \pi \sigma^3 /6$ is the network monomer volume with the diameter $\sigma=\sigma_\text{nn}=\sigma_\text{np}$. The expansion  \Eq~\eqref{eq:b2b3} is compared with the simulation data by fitting the parameters $B_2^\text{np}$ and $B_3^\text{nnp}$. The final best fits are depicted by the solid curves in \Fig~\ref{fig:fig2}a and are in very good agreement. The comparison implies the pronounced contribution of many-body ($B_3^\text{nnp}$) correlations, which are responsible for the non-monotonicity in the attractive and dense regimes.

The fitted $B_2^\text{np}$ and $B_3^\text{nnp}$ parameters can be found in \Tab~\ref{Tab:B2}.  We find that the second virial coefficients $B_2^\text{np}$ obtained from the fitting agree well with the values from the explicit relation  $B_2^\text{np}(\epsilon_\text{np})~=~\int_{0}^{\infty} \text{d}r 2 \pi~r^2[1~-~\exp(-\beta U_\text{LJ}^\text{np}(r,\epsilon_\text{np}))]$ for LJ particles, cf. \Fig~S2a in ESI. However, as shown in \Tab~\ref{Tab:B2} and \Fig~S2b in ESI, the third virial coefficient $B_3^\text{nnp}$ from the fitting deviates from the explicitly computed values of the LJ systems. This implies that as the polymer density increases many-body interactions, including the cross-linkers, play a major role, which is beyond the effect of a simple LJ liquid. In fact, the fitted $B_3^\text{nnp}$ values are always positive, \ie, the average many-body effect can be identified as on average a repulsive contribution. 

The data in  \Fig~\ref{fig:fig2}d is also well described by the virial form \Eq~\eqref{eq:b2b3}, where the solid lines agree with the simulation data. For this, we use \Eq~\eqref{eq:b2b3} with the same virial coefficients obtained from the result in \Fig~\ref{fig:fig2}a, and assume ${\phi_\text{n}}=\overline{\phi_\text{n}}$, which is in fact a good approximation particularly for low and high polymer densities.
The dependence of the partitioning on the network volume fraction can thus again be explained by a balance between the network--penetrant attraction and exclusion, which is particularly important for high volume fractions. 

\subsubsection{Diffusivity}

In \Fig~\ref{fig:fig2}b the penetrant diffusivity $D_\text{in}$ in the network is shown versus the polymer packing fraction $\phi_\text{n}$. Note that the diffusivity is rescaled by the diffusivity in the bulk $D_\text{0}$.  
The diffusivity is monotonically decreasing and tends to decay rapidly as the network volume fraction increases~\cite{HAUS1987,masaro1999physical, Amsden1998, Ghosh2014a, Lyd}.
  The dashed line depicts $\enat^{-\phi_\text{n}}$ for a simple exponential reference function. We furthermore compare the simulation results with the ``free-volume'' theory ~\cite{Yasuda1968,yasuda1969permeability2,yasuda1969permeability3,peppas1983solute,reinhart1984solute,lustig1988solute,Amsden1998},
\begin{eqnarray} 
D_\text{in}^{\rm fv}/D_\text{0} = b \exp \left[-c \left( \frac{\phi_\text{n}}{1-\phi_\text{n}} \right) \right].
\label{eq:gfree}
\end{eqnarray}
The solid lines show the fitting with the prefactor $b$ and the exponent $c$, which perform in an excellent fashion.
The fitting values of $b$ and $c$ are shown in \Tab~\ref{Tab:B2}. We note that $b$ decays exponentially as $\epsilon_\text{np}$ increases, while $c$ is rather independent (see \Figs~S7 and S8 in ESI for details). This is physically reasonable if we regard diffusion for large attractions as an activated process, in which the penetrants have to escape from locally bound states (`traps').  Therefore, here we present a semi-empirical scaling expression for the penetrant diffusivity,
\begin{eqnarray}\label{eq:D_scale}
D_\text{in}/D_0 \sim \enat^{-\beta\epsilon_\text{np}-c\left( \phi_\text{n} \over { 1 - \phi_\text{n} } \right)}. 
\end{eqnarray}

In \Fig~\ref{fig:fig2}e we confirm that $D_\text{in}(\epsilon_\text{np})$ indeed tends to exponentially decrease.
Hence, the Kramers' type scaling $D_\text{in} \propto \enat^{-\beta \epsilon_\text{np}}$ for the diffusion limited escape from a single attractive well~\cite{masaro1999physical} fits well, such that our prediction from \Eq~\eqref{eq:D_scale} holds. It is interesting that the energy barrier in the dense systems (\ie, the micro-roughness of the energy landscape) is simply described by $\epsilon_\text{np}$ and not by multiples of it, as we observed in the more ordered systems~\cite{kim2019prl}. Apparently, the random structure (\ie, polydispersity of the network) smoothens out the roughness significantly.
Note again that the overall mean attraction (\ie, the mean of the landscape in contrast to its roughness) is much higher than $\epsilon_\text{np}$, since we needed  $7 \epsilon_\text{np}$ to fit the partition ratio above. We remark that the scaling law \Eq~\eqref{eq:D_scale} has limitations since it does not behave well when $\phi_\text{n} \rightarrow 0$ where $D_{\rm in}/D_0$ should go to unity. However, this dilute limit with little influence on transport is not interesting anyway for applications and controlling the selectivity. We recall that in literature there are in fact various conventional scaling theories for the diffusivity~\cite{masaro1999physical, Amsden1998}. In \Fig~S7 in ESI, we present several appropriate scaling theories for the diffusivity compared to our simulation results, where \Eq~\eqref{eq:D_scale} performs the best throughout the range of $\phi_\text{n}$, including the dense regime.

It is interesting that the diffusivity is a simple monotonic function of $\phi_\text{n}$.  In fact, this result is very different from our previous finding for regular topologies, that is, membranes made of a fixed (static) fcc (face-centered-cubic) or simple-cubic lattice of LJ spheres~\cite{kim2019prl}. There, we found that the diffusivity is rather a complex function of the density of the membranes. We rationalized the effect by the roughness of a potential landscape, which for ordered potential wells on a regular lattice can be a very rapidly changing function of membrane density in certain density regions~\cite{kim2019prl}. 
But in the case here, the fluctuations and the polydispersity of the polymer network smoothen out the sharp density effects on the energy landscape and all diffusivities scale similarly exponentially, qualitatively almost independent of the  parameter $\epsilon_\text{np}$. 

\subsubsection{Permeability}

In  \Fig~\ref{fig:fig2}c we present the permeability $\mathcal{P}=~\mathcal{K} D_\text{in}$ versus the packing fraction.  The permeability varies by about 4 orders of magnitude in our parameter range.   Due to the generic behavior of the diffusion, the functional form of the permeability reflects essentially the one of the partition ratio $\mathcal K$, while the diffusivity only quantitatively scales the results. Hence, we find that for small interactions $\epsilon_\text{np}$, the permeability is monotonically decreasing with density, whereas for stronger interactions, it becomes a non-monotonic function of density. Therefore, as an important finding, the permeability can be maximized in our network model system. For the largest network--penetrant attraction,  the permeability is maximized at around $\phi_\text{n} \simeq 0.28$ by a factor of around 20 when compared to the bulk reference permeability $\mathcal P=D_0$ (the cross symbol). 

Having well-performing scaling laws for $\mathcal K$ and $D_{\rm in}$ from \Eqs~\eqref{eq:b2b3} and~\eqref{eq:D_scale}, we attempt to empirically construct also a scaling law for the permeability, via their product, eqn~(1),  
\begin{eqnarray}\label{eq:perm}
\mathcal{P} = \scriptstyle{ \exp \left[ -\beta\epsilon_\text{np} -c\left( \phi_\text{n} \over { 1 - \phi_\text{n} } \right) -2 B_2^\text{np} \frac{\phi_\text{n}}{v_0} - \frac{3}{2} B_3^\text{nnp} \left( \frac{\phi_\text{n}}{v_0} \right)^2 \right] },
\end{eqnarray}
comprising the attractive contribution as a function of the network--penetrant interaction $\epsilon_\text{np}$, and the exclusion contribution as a function of the packing fraction. The maximization of $\mathcal P$ can therefore be understood via \Eqs~\eqref{eq:b2b3} and~\eqref{eq:D_scale}.  The solid lines in \Fig~\ref{fig:fig2}c are the predictions from \Eq~\eqref{eq:perm} using the fit parameters determined already in panels a and b, showing very good agreement with the simulation results.

The permeability as a function of the network--penetrant interaction, $\mathcal{P}(\epsilon_\text{np})$, shown in \Fig~\ref{fig:fig2}f, is an increasing function from the global minimum at around $\beta\epsilon_\text{np} = 0.1$, which substantially depends on membrane density. Here, the selective tuning of $\mathcal{P}$ is mainly controlled by the penetrant's excluded volume. The prediction from the empirical scaling \Eq~\eqref{eq:perm} indeed agrees well with the simulation data, in particular, capturing the competition and cancellation between the exponentially growing partitioning and the exponentially decreasing diffusion. 

\subsection{Anti-correlations between $\mathcal K$ and $D_{\rm in}$ and tuning of the permselectivity}

\begin{figure}[t]
\centering
\includegraphics[width = 0.43\textwidth]{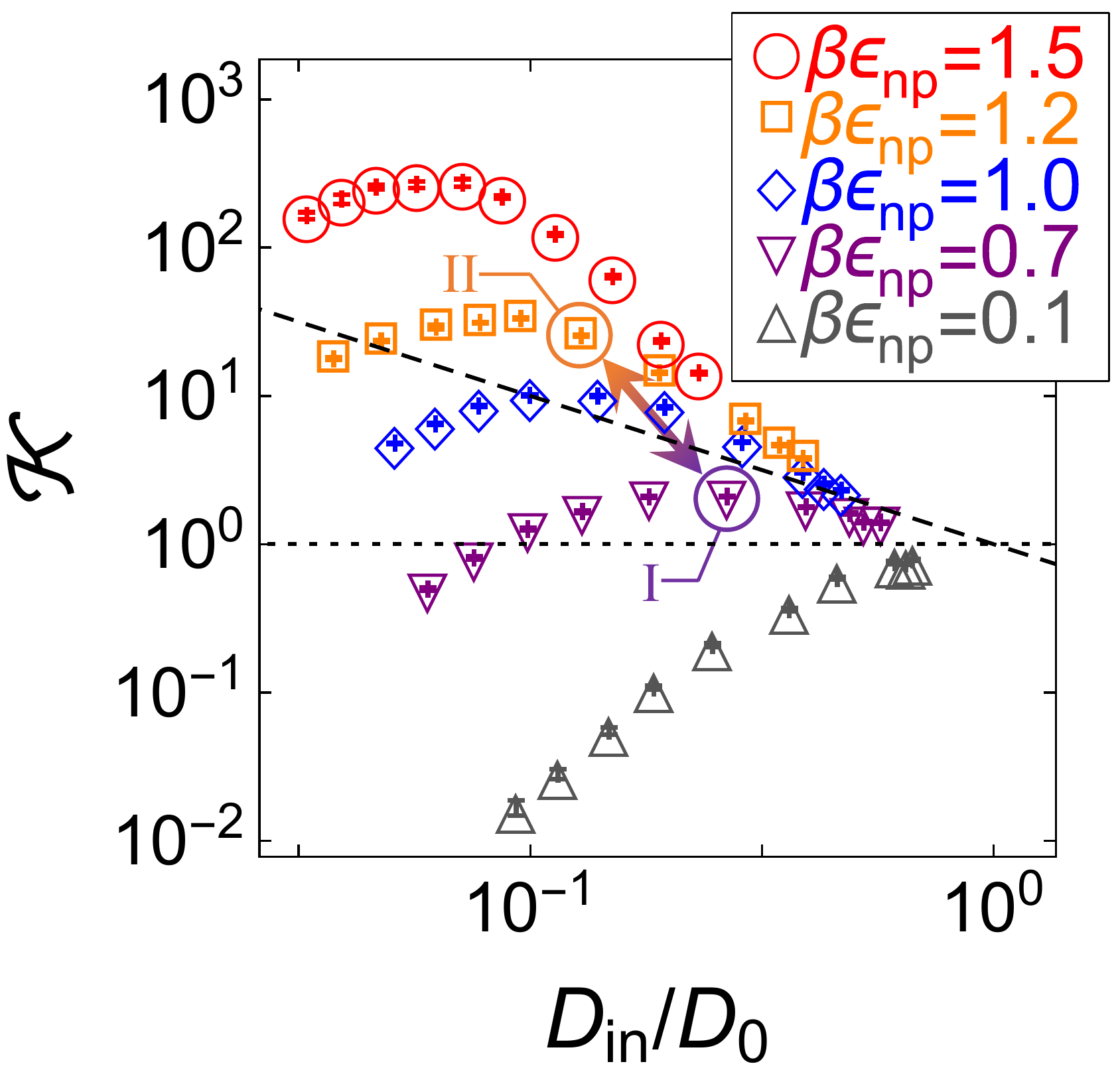}
\caption{
The partitioning--diffusion $\mathcal{K}$--$D_\text{in}/D_{0}$ correlation diagram. As depicted in the legend, symbols of the same color have the same network--penetrant interaction, but different polymer densities, \ie, the network--network interaction and hence the polymer volume fraction increases (for each color individually) from right (high diffusion) to the left (low diffusion). The black dashed line depicts the iso-permeability line $\mathcal{P}/D_{0}= \mathcal{K}D_\text{in}/D_{0}=1$ (reference bulk permeability), where the actions of $\mathcal{K}$ and $D_\text{in}$ on the membrane permeability exactly cancel each other. The arrow connects two states II $(\beta\epsilon_\text{np}=1.2)$ and I $(\beta\epsilon_\text{np}=0.7)$ at packing fraction $\phi_\text{n} \simeq 0.17$, featuring the selectivity ratio $\alpha_\text{II,I} \equiv \mathcal{P}_\text{II}/\mathcal{P}_\text{I} \simeq 6$, see text.
 }
\label{fig:fig3}
\end{figure}

The diagram in \Fig~\ref{fig:fig3} plots partitioning $\mathcal K$ versus diffusivity $D_{\rm in}$ and thus presents a landscape visualizing how they are correlated, \ie, a partitioning--diffusivity correlation diagram. The plot shows a wide landscape of the permeability spanning over several orders of magnitude. The black dashed line depicts the iso-permeability line of the bulk permeability $\mathcal{P}/D_{0}=1$, where the two contributions exactly cancel out.  The data in \Fig~\ref{fig:fig3} at low and intermediate polymer densities lead to final permeabilities close to the iso-permeability line, hence exhibiting clear anti-correlations and cancellations. Such a cancellation was also observed, even massively leading to more qualitative changes, in membranes constructed by static regular obstacles~\cite{kim2019prl}. This can be understood by going back to our scaling law, \Eq~\eqref{eq:perm}. The attraction between monomers and penetrants increases the uptake of penetrants in the membrane roughly exponentially. However, at the same time the attraction enhances the microscopic roughness and deepens local traps, thereby impeding the thermally activated escape, which in turn also leads to an exponential decrease of diffusion. In many regimes, these two effects cancel out, but the exact behavior depends on the details of the variation of the energy landscape~\cite{kim2019prl}. This can be harvested to tune and optimize the selectivity of a polymer membrane. However, in contrast to the ordered membranes~\cite{kim2019prl}, this work indicates that the diffusivity in polydisperse networks only rescales the permeability, while the functional form is dictated by the partitioning behavior.

Hence, the diagram in \Fig~\ref{fig:fig3} presents non-trivial pathways of the permeability $\mathcal{P}$ along the two variable parameters, density and penetrant--network attraction.  It clearly shows how the permeability can be tuned substantially over several orders of magnitude already by a relatively small material parameter space. With this, a significant selective permeability (permselectivity) can be demonstrated depending on the interaction parameter $\epsilon_\text{np}$ (which in reality is different for various chemically specific penetrants). For instance, defined as $\alpha_\text{II,I} \equiv~\mathcal{P}_\text{II}/\mathcal{P}_\text{I}$~\cite{freeman1999basis}, the selectivity for the states II $(\beta\epsilon_\text{np}=1.2)$ and I $(\beta\epsilon_\text{np}=0.7)$ depicted by the arrow in \Fig~\ref{fig:fig3} at a packing fraction $\phi_\text{n} \simeq 0.17$ amounts to $\alpha_\text{II,I}\approx6$, which is large. Hence, a small difference of interactions of half a $\kT$ results already in a permeability ratio of almost one order of magnitude.  

\section{Conclusion}

We presented extensive (implicit-solvent) coarse-grained simulations and scaling theories for penetrant transport through semi-flexible, cross-linked, and polydisperse polymer networks with a focus on the linear-response permeability, calculated by the equilibrium partitioning and diffusion of the penetrants inside the network.  The permeability has been found to be largely tunable by varying the polymer network density and the microscopic interactions between the network and the diffusive penetrants.  In particular, significant maximization and minimization of the permeability were found, fine-tuned by the solvent quality and the penetrant--network interactions.  The results were rationalized by scaling theories which include a virial expansion with two-body attractions and many-body exclusion effects for the partitioning, and a combination of the free-volume and Kramers' escape scaling laws for the diffusivity. The presented laws, despite their simplicity, capture salient features of the system, showing good agreement with the simulation results.

The penetrant diffusivity turned out to be rather a smooth function of the network density, implying substantial effects of the fluctuation and randomness of the polymer network. The polydisperse nature of the network averages out the roughness of the energy landscape, which was more pronounced and sensitive to parameter changes in highly ordered, lattice-based and static membrane systems~\cite{kim2019prl}. Nevertheless, the permeability revealed a rather intricate, non-monotonic behavior over several orders of magnitude, originating from the complex nature of the partitioning, while quantitatively and substantially modified by the anti-correlated and canceling contributions of the diffusion. As a consequence, only small changes of interactions, \eg, by half a $\kT$ can already modify the selectivity of the membrane by a factor of 6. 
Our study provides a further step in the fundamental understanding and development of a minimal theory to characterize better the permeability in flexible and fluctuating polymer-based membrane systems.

\section*{Conflicts of interest} There are no conflicts to declare.

\section*{Acknowledgements}
The authors thank Matthias Ballauff, Benjamin Rotenberg, Arturo Moncho-Jord{\'a} and Changbong Hyeon for fruitful discussions. This project has received funding from the European Research Council (ERC) under the European Union's Horizon 2020 research and innovation programme (grant agreement No.\ 646659). W.K.K. acknowledges the support by a KIAS Individual Grant (CG076001) at Korea Institute for Advanced Study. M.K. acknowledges the financial support from the Slovenian Research Agency (research core funding No.\ P1-0055). The simulations were performed with resources provided by the North-German Supercomputing Alliance (HLRN). We thank Center for Advanced Computation at the Korea Institute for Advanced Study for providing computing resources for this work.



\end{document}